\input{aipcheck}

\documentclass[
    ,final            % use final for the camera ready runs
  ]
  {aipproc}

\layoutstyle{8x11double}

\renewcommand\XFMtitleblock{%
  \XFMtitle
  \let\XFMoldpar\par
  \def\par{\XFMoldpar\def\par{\space 
             (on behalf of the VERITAS Collaboration)\XFMoldpar}}%
   \XFMauthors
   \let\par\XFMoldpar
   \XFMaddresses
   \XFMabstract
   \vspace{5pt}%
   \XFMkeywords
   \XFMclassification
 }

\begin{document}

\title{Observation of Galactic Gamma-ray Sources with VERITAS}

\classification{95.85.Pw, 97.10.Gz, 97.60.Gb, 97.60.Lf, 97.80.Jp, 98.70.Rz}
              
\keywords      {gamma rays: observations}

\author{G.Maier}{
  address={Department of Physics, McGill University, H3A 2T8 Montreal, QC, Canada \\ maierg@physics.mcgill.ca \\ see http:\/\/veritas.sao.arizona.edu\/  for a list of members } }

\begin{abstract}
We report on VERITAS observations at energies above 200 GeV of known or potential
galactic $\gamma$-ray sources.
The observed objects comprise pulsars, pulsar wind nebulae,
high-mass X-ray binaries and gamma-ray sources with unknown counterparts in other wavelengths.
Among the highlights are the observation of variable gamma-ray emission from the X-ray binary
LS I +61 303 and the detection of MGRO J1908+06/HESS J1908+063, an extended gamma-ray source which
could not be associated with any obvious counterpart at lower energies.
\end{abstract}

\maketitle

%%%%%%%%%%%%%%%%%%%%%%%%%%%%%%%%%%%%%%%%%%%%
%% MAINMATTER
%%%%%%%%%%%%%%%%%%%%%%%%%%%%%%%%%%%%%%%%%%%%

Discoveries by the current generation of instruments have revealed a rich
variety of galactic sources.
The different classes of galactic sources range from 
supernova remnants, pulsars and pulsar wind nebulae to X-ray and Wolf-Rayet
binaries.
The gamma-ray emission is often interpreted as inverse Compton scattering of 
low-energy photons by relativistic electrons.
Another possibility
is pion production in hadronic interactions, 
which links directly to the problem of galactic cosmic ray acceleration.
Measurements of the spatial extension, spectral energy distribution,
and temporal variability of gamma-ray emission from galactic sources
with the VERITAS array as presented in the following 
might reveal the nature and origin of high-energy radiation in 
our Galaxy.
Results from VERITAS observations of supernova remnants can be found elsewhere
\cite{Hummensky2008}.

\section{VERITAS}

VERITAS is an array of four imaging Cherenkov
telescopes located at the Fred Lawrence Whipple
Observatory in southern Arizona.
It combines
a large effective area ($>8 \times 10^4$ m$^2$) over
a wide energy range (100 GeV to 30 TeV) with
good energy (15-20\%) and angular resolution
($\approx 0.1^{\mathrm{o}}$).
The high sensitivity of VERITAS allows the detection
of sources with a flux of 1\% of the Crab
Nebula in under 50 hours of observations
\cite{Holder2008}.

\section{Pulsar Wind Nebula}

\begin{table}
\begin{tabular}{lccccc}
\hline
  \tablehead{1}{l}{b}{Object \\ }
  & \tablehead{1}{c}{b}{log$\mathbf{_{10}}$ $\dot \mathbf{E} \mathbf{/d^2}$ \\ $\mathbf{[erg/s/kpc^2]}$ }
  & \tablehead{1}{c}{b}{T$\mathbf{_{obs}}$\tablenote{dead time corrected observation time} \\ $[$hours$]$ }
  & \tablehead{1}{c}{b}{$<$Z$>$\tablenote{mean zenith angle of observations} \\ $[$deg$]$ } 
  & \tablehead{1}{c}{b}{Significance \\ $[\sigma]$} 
  & \tablehead{1}{c}{b}{Flux ($\mathbf{>300 GeV}$)\tablenote{assuming a point-like source} \\ $[$\% Crab$]$ } \\
\hline
Crab Nebula & 38.1 & 11 & 15 & 100 & 100 \\
PSRJ0205+6449 & 36.4 & 12.8 & 35.2 & 1.1 & $<$2.3 \\
PSRJ0631+1036 & 33.6 & 13.0 & 24.5 & 0.3 & $<$1.3 \\
PSRJ0633+1746 & 36.1 & 13.3 & 17.6 & -0.1 & $<$1.0  \\
PSRJB0656+14 & 35.6 & 9.4 & 22.4 & -1.8 & $<$0.2  \\
PSRJ1740+1000 & 35.1 & 10.5 & 24.6 & 0.2 & $<$1.0 \\
\hline
\end{tabular}
\caption{\label{tabl:PWN} Results from VERITAS observations at energies above 300 GeV of 
selected pulsar wind nebula.}
\label{tab:a}
\end{table}

Most galactic sources of TeV gamma-rays have been identified as pulsar wind nebula.
These are thought to be young pulsar which lose their rotational energy by relativistic winds
in which particles can be accelerated to very-high energies.
VERITAS observed a selection of pulsars with large energy loss 
($\dot \mathrm{E} \mathrm{/d^2} > 10^{35}$ ergs/s/kpc$^2$) for around 10 hours each at
zenith angles ranging from 15 to 35$^{\mathrm{o}}$, see Table \ref{tabl:PWN}.

The Crab Nebula is one of the strongest known gamma-ray sources. 
It has been extensively observed with VERITAS and is typically detected with 
a significance larger than 30$\sigma/\sqrt{\mathrm{hour}}$.
The observations of the Crab Nebula are valuable both for astrophysical \cite{Celik:2007} and technical 
studies \cite{Maier:2007}.

None of the observed PWN, with the exception of the Crab Nebula,
has been detected as emitter of high-energy gamma-rays, 
99\% upper flux limits are in the range from 0.2 to 2.3\% of the flux of the Crab Nebula 
(Table \ref{tabl:PWN}).
The absence of any detected TeV flux from these 
objects is most likely explained by the relatively short 
VERITAS exposures.
Some of the most promising objects in this sample will be re-observed in the future.
Alternative possibilities include the fact that the moderately 
old PWN in the sample may no longer be confined by 
their parent SNRs.
In this case, the confinement is provided rather by the ISM,
which may be of rather low density. 
For more details see \cite{Aliu2008}.

\section{Search for pulsed $\gamma$-ray emission from Geminga}

%\begin{figure}
%  \includegraphics[height=.3\textheight]{Geminga_ID32_cut-NImage34_RM_radec}
%  \caption{Picture to fixed height}
%\end{figure}

\begin{figure}
  \includegraphics[width=0.9\linewidth]{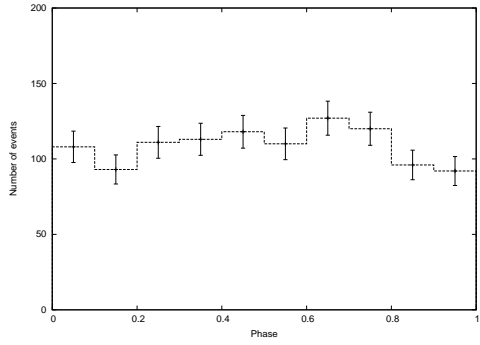}
  \caption{\label{fig:Geminga}Phase histogram for Geminga from VERITAS gamma-ray observations
  for energies above 300 GeV.}
\end{figure}

The measurement of pulsed emission at energies above several hundred GeV 
sets constrains to emission models of pulsars.
Geminga (PSR J0633+1746) is one of the strongest EGRET sources ($>$100 MeV) for steady and 
pulsed emission with a period of \mbox{237 ms} \cite{Halpern:1992} and has been associated
with a gamma-ray source by MILAGRO \cite{Abdo2007}.
Hard X-ray and soft $\gamma$-ray observations point to electron acceleration up to energies 
of $10^{14}$ eV \cite{Caraveo2003}.
Upper limits between 5 and 12\% of the steady flux of the Crab Nebula has been reported by
HEGRA \cite{Aharonian:1999} and Whipple \cite{Akerlof:1993}.
Figure \ref{fig:Geminga} shows the phase histogram for Geminga from VERITAS gamma-ray observations.
No pulsed emission at energies above 300 GeV has been detected during 13 hours of observations.
An upper flux limit of $1\times10^{-8}$ $\gamma$/m$^2$/s (about 0.5\% of the steady flux of the Crab Nebula)
has been derived \cite{Kieda2008}.

\section{The gamma-ray binary LS I +61 303}

\begin{figure}
   \includegraphics[width=1.0\linewidth]{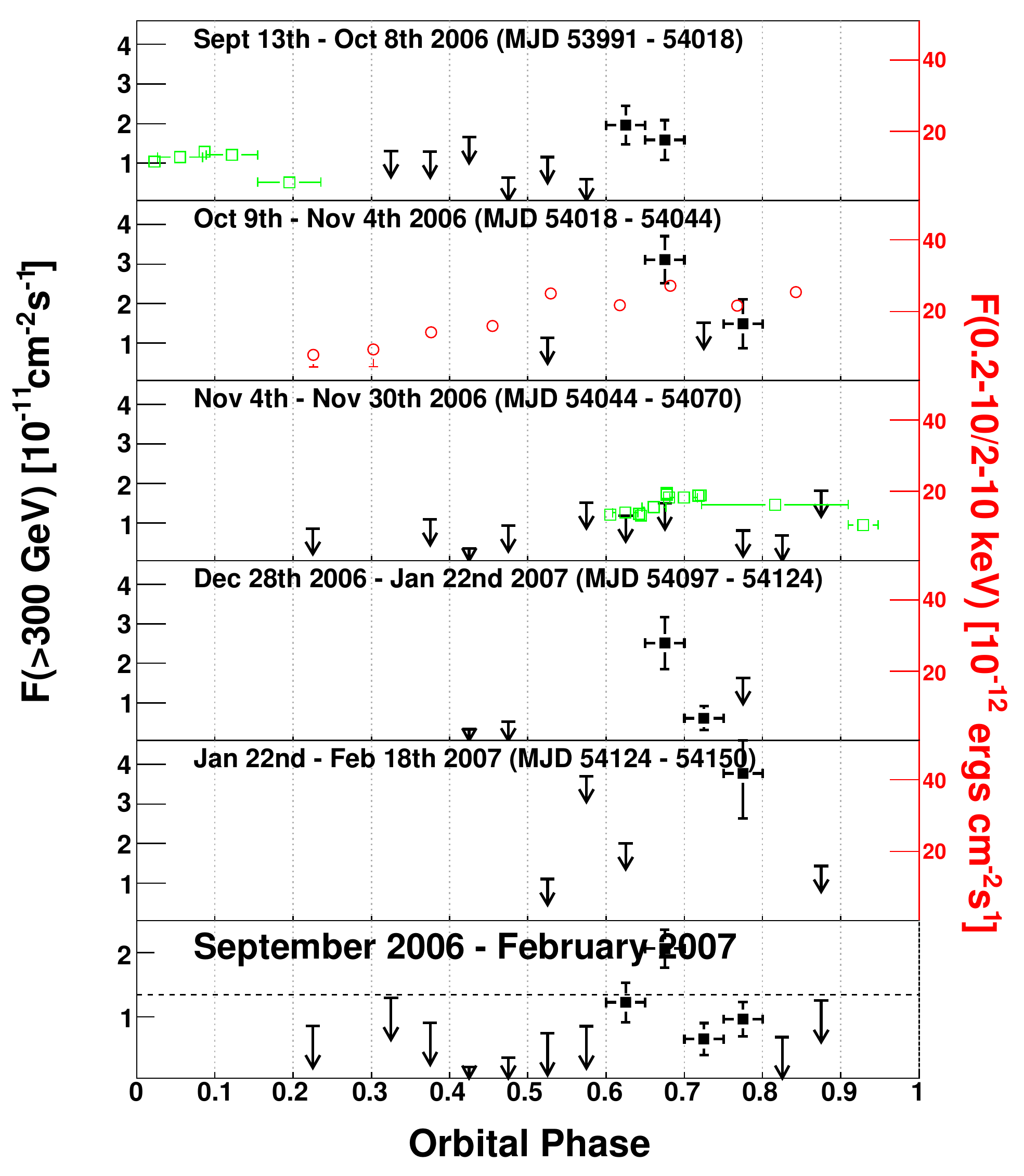}
   \caption{\label{fig:LSI2006}
   Average fluxes or upper flux limits per orbital phase bin for
   $\gamma$-rays with energies above 300 GeV from the direction
   of LS I +61 303 as a function of orbital phase for the observation
   period September 2006 to February 2007.
   The bottom panel shows the results averaged over the whole dataset,
   the upper panels show the results for the individual plots.
   Upper flux limits (95\% probability \cite{Helene1983}) are shown
   for data points with significances less than 2$\sigma$ (significance calculation 
   after \cite{Li1983}).
   The hard X-ray fluxes as observed by RXTE (2-10 keV, red circles) and 
   Swift (0.2-10 keV, green squares) are indicated with open markers.}
\end{figure}

\begin{figure}
   \includegraphics[width=1.0\linewidth]{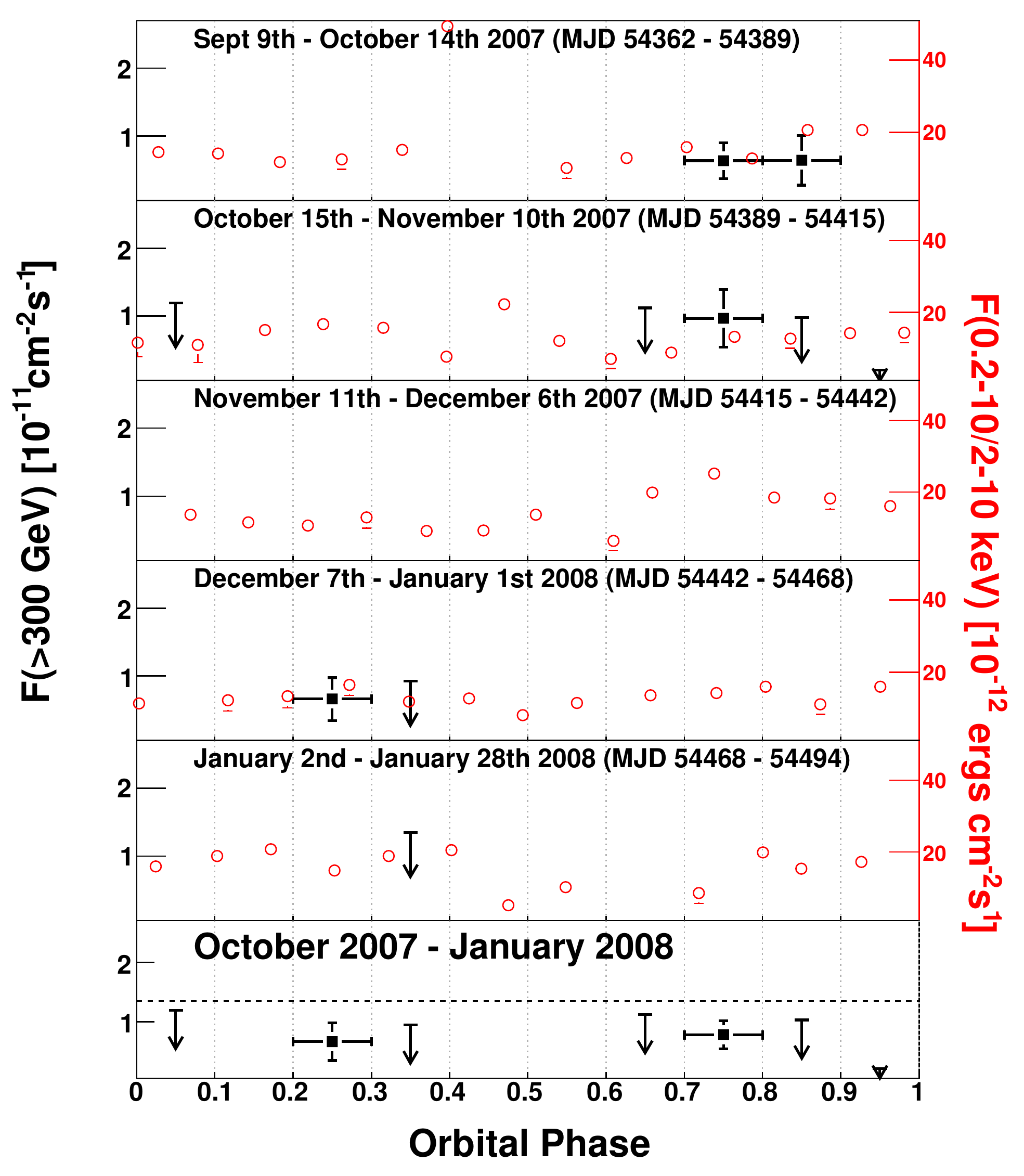}
   \caption{\label{fig:LSI2007}
   Same as Figure \ref{fig:LSI2006}, but for observation period September 2007 to January 2008.}
\end{figure}

The high-mass X-ray binary system LS I +61 303
is one of only three X-ray binaries detected in high-energy
gamma rays \cite{Aharonian05,Aharonian06,Albert06,Acciari2008}.
It consists of a massive
Be-type star surrounded by a dense circumstellar
disk and a compact object (neutron star or
black hole).
Optical observations show that the
compact object orbits the star every 26.5 days \cite{Gregory02}
on a close orbit, characterized by a semi-major axis
of a few stellar radii only.
The unknown nature of the compact object and
the uncertainty in the geometry of the system allows
the existence of at least two possible models
for the origin of the high-energy emission from
LS I +61 303.
The first class of models explains
the emission of 
gamma-rays through the production of
non-thermal particles in an accretion powered relativistic
jet (microquasar model).
In the
second class of models, particles are accelerated in
the shock created by the collision of a relativistic
pulsar wind with the wind of the companion star.
In both cases, the 
gamma-ray emission is interpreted
as inverse Compton scattering of stellar radiation
by high-energy electrons. Alternative models
of 
gamma-ray production include IC e$\pm$ pair cascades in
the field of massive stars \cite{Bednarek06}
or $\pi^{0}$ production and
decay in hadronic interactions.
\cite{Romero05}.
It is expected that the dense and anisotropic photon fields
in LS I +61 303 lead to energy dependent modulation of the
low- and high-energy gamma-ray fluxes \cite{Dubus2008}.

LS I +61 303 has been observed with VERITAS for 45 hours between
September 2006 and February 2007 (Figure \ref{fig:LSI2006}) and for 12 hours between September 2007 and 
January 2008 (Figure \ref{fig:LSI2007}).
Most of the 2006-2007 data was taken during the construction of VERITAS
with two or three telescopes only.
The 2007-2008 data was taken partly under moonlight conditions.

The detected flux from the direction of LS I +61 303
is measured to be strongly variable;
the maximum flux (corresponding to about 10\% of the flux of the Crab Nebula)
is always found at approximately apastron.
This suggests that the gamma-ray production
region is close to or inside the binary system and
that the tight orbit, combined with the dense stellar
wind of the Be-star, produces a continuously
changing environment for particle acceleration and
absorption.
It is unclear at this point if there is a strong correlation
between the X-ray and gamma-ray emission from LS I +61 303 \cite{Smith2008};
this will be studied in future observations of VERITAS
together with observations from various X-ray satellites.

\section{Unidentified Galactic Sources}

Among the major discoveries in high-energy gamma-ray astronomy
in the last few years are the detection of previously unknown
classes of sources of high-energy gamma-rays (e.~g.~Wolf-Rayet binaries),
transient gamma-ray emitter (e.~g.~X-ray binaries), and
of sources with no obvious counterparts in other wavelengths.
Two unidentified sources are the AGILE transient source
AGILE 2021+4024 \cite{Longo2008} and the MILAGRO unidentified source
MGRO J1908+06 \cite{Abdo2007}

\subsection{AGILE 2021+4024}

\begin{figure}
  \includegraphics[width=0.8\linewidth]{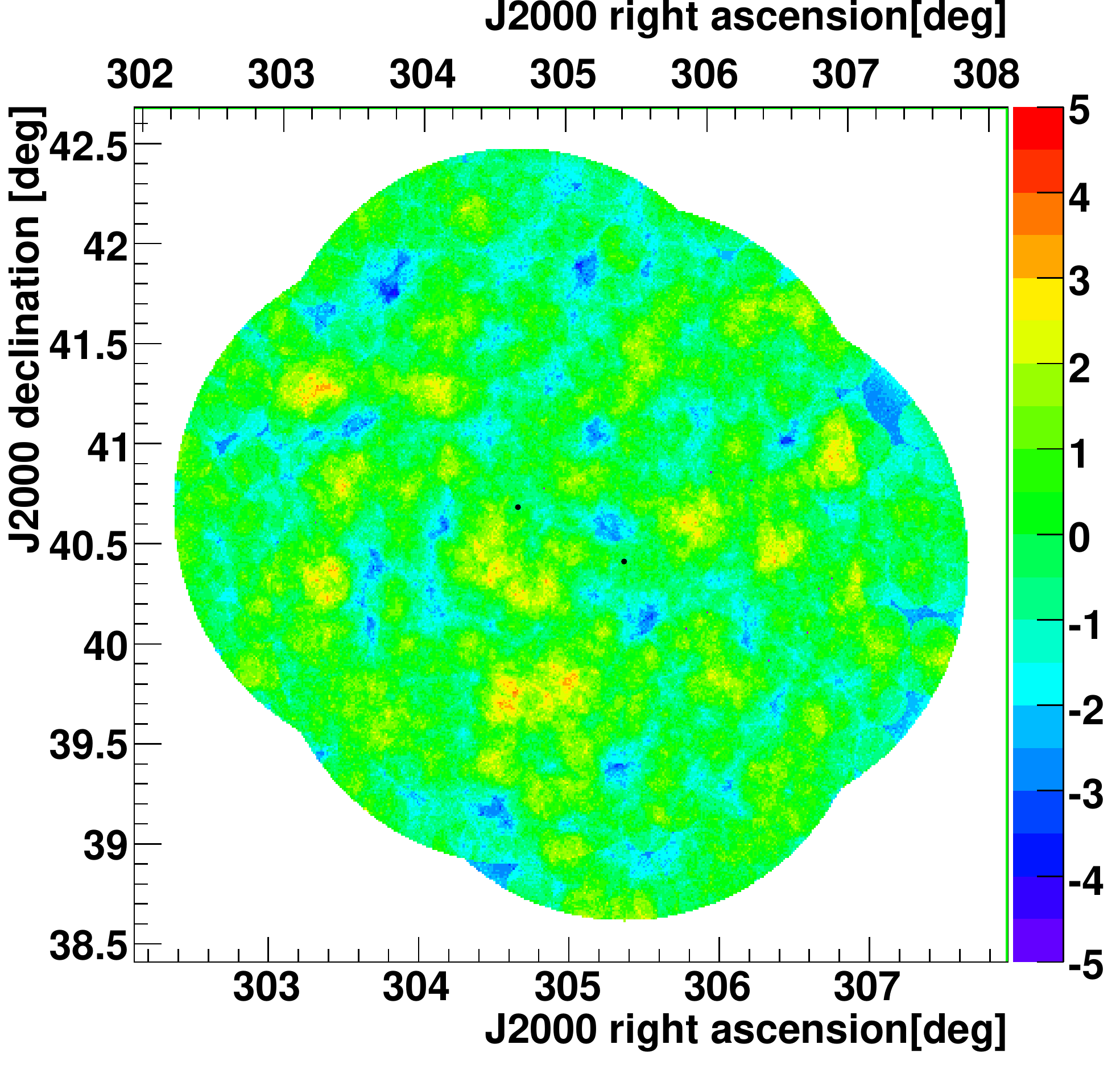}
  \caption{\label{fig:Agile} Significance map derived from 7 hours
  of VERITAS observations of the region around
  AGILE 2021+4024 in equatorial J2000 coordinates.}
\end{figure}

The AGILE satellite detected significant and variable gamma-ray 
emission above 100 MeV from a source positionally consistent with the 
EGRET source 3EG J2020+4017 during observations in
April 2008. The variable source rebrightend twice in May and June 2008 
\cite{Giuliani2008, Chen2008}. 
VERITAS observed the region around AGILE 2021+4024 during the nights
from April 29th to May 6th 2008 for about 7 hours.
The sky plot of the region around AGILE 2021+4024 in gamma rays
with energies above 300 GeV does not reveal any significant emission 
(see Figure \ref{fig:Agile}),
an upper limit for gamma rays with energies above 300 GeV
at the 99\% confidence level of about
2\% of the steady flux of the Crab Nebula has been derived
for the reported position of AGILE 2021+4024 from 
these VERITAS observations.

\subsection{MGRO J1908+06}

\begin{figure}
  \includegraphics[width=0.8\linewidth]{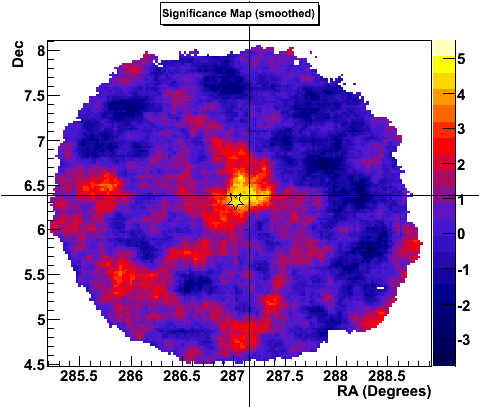}
  \caption{\label{fig:MGRO} Significance map derived from 22 hours of VERITAS observations of
  the region around
  MGRO J1908+06 in equatorial J2000 coordinates.}
\end{figure}

The unidentified gamma-ray source MGRO J1908+06 has been detected in a survey of the northern hemisphere sky
for sources of TeV gamma-rays with the MILAGRO Gamma-Ray Observatory \cite{Abdo2007}.
The reported extension of the sources is $<2.6^{\mathrm{o}}$ with a flux of about 80\% of the Crab Nebula flux
at a median energy of 20 TeV.
The position of HESS J1908+063, a high-energy gamma-ray source detected by the H.E.S.S. collaboration \cite{Djannati-Atai2007}
is compatible within errors with the position of MGRO J1908+06. 
The extension of HESS J1908+063 is 0.21$^{\mathrm{o}}$ at
energies above 300 GeV with a flux of 14\% of the Crab Nebula flux at 1 TeV.
Counterpart of this unidentified source might be the radio-bright supernova remnant SNR G040.5-00.5
at an estimated distance of about 5.3 kpc. 
VERITAS observations of the region around MGRO J1908+06 reveals an extended gamma-ray source with an extension and a position in agreement
with HESS J1908+063, see Figure \ref{fig:MGRO}.
For more details see \cite{Ward2008}.

\section{Conclusions}

The first year of VERITAS observations of galactic gamma-ray sources has been successfully completed.
Among the detected sources are the Crab Nebula, the X-ray binary LS I +61 303, the supernova remnants
IC 443 and Cas A \cite{Hummensky2008} and the unidentified source MGRO J1908+06/HESS J1908+063.
The excellent performance of VERITAS demonstrate the great potential
of the system for the discovery of new gamma-ray sources and
for detailed morphological and spectral studies.

\begin{theacknowledgments}
This research is supported by grants from the U.S.
Department of Energy, the U.S. National Science
Foundation, and the Smithsonian Institution, by
NSERC in Canada, by PPARC in the UK and by
Science Foundation Ireland.
\end{theacknowledgments}

\end{document}